\documentclass{article}

\usepackage{amsthm}

\usepackage{PRIMEarxiv}

\usepackage[utf8]{inputenc} 
\usepackage[T1]{fontenc}    
\usepackage{hyperref}       
\usepackage{url}            
\usepackage{booktabs}       
\usepackage{amsfonts}       
\usepackage{nicefrac}       
\usepackage{microtype}      
\usepackage{lipsum}
\usepackage{fancyhdr}       
\usepackage{graphicx}   
\usepackage{graphicx} 
\usepackage{booktabs}
\usepackage{todonotes}
\usepackage{doi}
\usepackage{float}
\usepackage{algorithm}
\usepackage{algorithmic}
\usepackage{multirow}
\usepackage{amsmath}
\newtheorem{theorem}{Theorem}
\newtheorem{definition}{Definition}
\graphicspath{{media/}}    

\title{Enhanced Renewable Energy Forecasting and Operations through Probabilistic Forecast Aggregation}
\author{\large Alireza Moradi, Mathieu Tanneau, Reza Zandehshahvar, Pascal Van Hentenryck\\\\
NSF Artificial Intelligence Institute for Advances in Optimization,\\
H. Milton Stewart School of Industrial and System Engineering,\\
Georgia Institute of Technology, Atlanta, GA, USA}

\begin{document}
\maketitle

\begin{abstract}
{Accurate and reliable forecasting of renewable energy generation is crucial for the efficient integration of renewable sources into the power grid. In particular, probabilistic forecasts are becoming essential for managing the intrinsic variability and uncertainty of renewable energy production, especially wind and solar generation. This paper considers the setting where probabilistic forecasts are provided for individual renewable energy sites using, e.g., quantile regression models, but without any correlation information between sites. This setting is common if, e.g., such forecasts are provided by each individual site, or by multiple vendors. However, to effectively manage a fleet of renewable generators, it is necessary to aggregate these individual forecasts to the fleet level, while ensuring that the aggregated probabilistic forecast is statistically consistent and reliable. To address this challenge, this paper presents the integrated use of Copula and Monte-Carlo methods to aggregate individual probabilistic forecasts into a statistically calibrated, probabilistic forecast at the fleet level. The proposed framework is validated using synthetic data from several large-scale systems in the United States. This work has important implications for grid operators and energy planners, providing them with better tools to manage the variability and uncertainty inherent in renewable energy production.}
\end{abstract}


\section{Introduction}
\label{sec:introduction}

As modern power grids integrate ever-higher levels of renewable energy generation, especially wind and solar generation, transmission systems operators (TSOs) are faced with increasing levels of operational uncertainty.
This operational challenge motivates the development of probabilistic forecasting methods,
which not only provide point estimates of, e.g., renewable energy production,
but can also quantity the associated uncertainty \cite{Zhang2025_ProbabilisticForecasting}.
Accordingly, several forecast vendors now offer probabilistic forecasts of power output for individual renewable generators.

This paper considers the setting where a TSO operates a power grid comprising $N$ renewable generators,
and has access to site-level probabilistic forecasts provided by a third-party vendor.
However, grid operators make decisions based on aggregate, fleet-level generation,
which raises the question of how to combine individual, site-level probabilistic forecasts,
into an aggregate, fleet-level probabilistic forecast.
It is important to note that, unlike point forecasts, site-level probabilistic forecasts cannot simply be summed.
Furthermore, the paper assumes that TSOs do not have access to a fleet-level probabilistic
forecast product, and have limited computational resources, thus preventing them
from developing their own forecasting tools from scratch.
Indeed, state-of-the-art forecasting tools typically require large amounts of climate data,
and extensive computational resources \cite{GraphCast}.

\paragraph{Related Works}
\label{sec:intro:literature_review}

Time series forecasting in power systems has been extensively studied, with methods evolving from traditional statistical models such as ARIMA \cite{miswan2016arima,alberg2018short} to machine learning approaches like support vector regression \cite{zahid2019electricity}, quantile regression forests \cite{8272993,he2020day}, and advanced deep learning techniques \cite{8863951,amarasinghe2017deep}. The increasing integration of renewable energy resources (RES) has shifted the focus toward high-dimensional probabilistic forecasting, which captures interdependencies between different targets and provides uncertainty-aware solutions for improved operations.

A key challenge in power system forecasting is the need for coherent forecasts across different levels of aggregation, including site, zone, and system levels. Ensuring that aggregated probabilistic forecasts remain valid and maintain consistency across hierarchical levels is essential for reliable grid operations.

Several methodologies have been developed to address the probabilistic forecast aggregation problem.
In \cite{taieb2021hierarchical}, the authors propose a bottom-up procedure, where marginal predictive distributions are used to estimate a joint distribution via an empirical copula, ensuring coherence in the aggregated forecasts.
Similarly, \cite{sun2019conditional} introduces a clustering-based approach for wind farms, reducing computational costs before employing a copula-based method to capture inter-cluster dependencies. 
Additionally, \cite{taieb2017coherent} models the dependency structure using a copula framework, enabling the computation of a coherent summation distribution from marginal predictive distributions.

\paragraph{Contributions}
This paper makes the following contributions.
First, it proposes a data-driven methodology to aggregate site-level probabilistic forecasts using copula functions.
The proposed methodology only requires access to historical site-level forecasts and actual generation.
Second, the paper presents numerical experiments on a large-scale system with several hundreds of solar sites.
The results demonstrate that the proposed approach outperforms two baselines that do not capture correlations,
and highlight the adverse effect of having poor site-level forecasts.

\section{Methodology}
\label{sec:methodology}

This section presents the mathematical formulation of the probabilistic aggregation in time series forecasting in power systems. 
Given the marginal probabilistic forecasts at site levels, the objective of this paper is to provide an accurate probabilistic forecast at the fleet level.
 
\subsection{Problem Setting}

Consider a system with $\mathcal{N} \, {=} \, \{1, ..., N\}$ renewable generators, e.g., solar farms, and a historical time window indexed by $t \, {\in} \, \mathcal{T} \, {=} \, \{1, ..., T\}$.
Let $x_{i, t}$ denote the actual power generation of site $i$ at time $t$, where $i \in \mathcal{N}$ denotes the site, and $t \in \mathcal{T}$ denotes time.
Let $\mathbf{x}_{., t} = [x_{1, t}, \dots, x_{N, t}]$ be the vector of $N$ actual generations at time $t$,
and define
\begin{align}
\mathbf{X}_{., \mathcal{T}}=
    \begin{bmatrix}
        \vert & \dots & \vert \\
        \mathbf{x}_{., 1} & \dots & \mathbf{x}_{., T} \\
        \vert & \dots & \vert 
    \end{bmatrix}
    \in
    \mathbb{R}^{N \times T}
    .
\end{align}
Next, let $\hat{F}_{i, t}$ denote the probabilistic forecast
for site $i \in \mathcal{N}$ for time $t \in \mathcal{T}$.
Namely, $\hat{F}_{i, t}$ denotes a univariate probability distribution,
also referred to as the \emph{marginal} forecast for site $i$ at time $t$.
Recall that probabilistic forecasts are produced at the site-level, i.e.,
no correlation information between sites is available.
This is the case in practice if, e.g., forecasts for two different 
sites are produced by two different vendors.

Finally, let $x_{0, t}$ denote the total power generation at the fleet level at time $t$,
i.e., $\forall t \in \mathcal{T}, x_{0, t} = \sum_{i=1}^{N}x_{i,t}$.
Given $\tau > T$ and marginal forecasts $(\hat{F}_{i, \tau})_{i \in \mathcal{N}}$,
the goal of the paper is to produce a probabilistic forecast for the total generation,
denoted by $\hat{F}_{0, \tau}$.
This setting represents, e.g., day-ahead operations, wherein future generation is obviously unknown,
site-level forecasts are available, and one seeks to provide operators with an accurate fleet-level forecast.

\subsection{Learning the Correlation Structure with Copula}
\label{sec:methodology:copula}

To address the above problem, the paper proposes to learn the correlation structure between sites,
by estimating a copula function from historical actuals and marginal forecasts.
Once learned, the copula allows to capture the dependencies
across marginal forecasts, from which accurate aggregated forecasts can be obtained.

\begin{definition}[Copula]
    Let $d \in \mathbb{Z}_{+}$.
    A function $\mathbf{C}: [0, 1]^{d} \rightarrow [0, 1]$ is a \emph{copula} if it is the cumulative distribution function (CDF)
    of a $d$-dimensional random variable $U$ whose marginal distributions are all uniform distributions
    $\mathcal{U}[0, 1]$.
\end{definition}

\begin{theorem}[Sklar's Theorem]
    Let $\mathbf{Y}$ be a $d$-dimensional random variable with joint CDF $\mathbf{F}$ and marginal CDFs $(F_{i})_{i \in \{1, \dots, d\}}$.
    There exists a copula function $\mathbf{C}: [0, 1]^d \rightarrow [0, 1]$ that connects the marginals to the joint distribution as:
    \begin{align}
        \mathbf{F}(y_1, \dots, y_d) = \mathbf{C}(F_{1}(y_1), \dots, F_{d}(y_d)).
    \end{align}
\end{theorem}

Sklar's theorem allows to model a joint distribution as the product of marginal distributions and a copula function that capture the correlation structure.
The paper uses the popular Gaussian copula, which has been successfully used in probabilistic forecasting, see, e.g., \cite{Zhang2025_ProbabilisticForecasting}.

The first step of the proposed methodology thus consists in estimating a copula
function, given historical actuals $(x_{i,t})_{i \in \mathcal{N}, t \in \mathcal{T}}$
and marginal forecasts $(\hat{F}_{i,t})_{i \in \mathcal{N}, t \in \mathcal{T}}$.
Let $i, t \in \mathcal{N} \times \mathcal{T}$, and define
\begin{align}
    \hat{y}_{i, t} = \hat{F}_{i, t}(x_{i, t}).
\end{align}
Note that $\hat{y}_{i, t}$ is a random variable with uniform distribution $\mathcal{U}{U}[0, 1]$.
Next, let $\Phi$ denote the CDF of a standard normal distribution, and define
\begin{align}
    \hat{\mathbf{Z}}
    &=
    \begin{pmatrix}
        \hat{z}_{1,1} & \dots & \hat{z}_{1, T}\\
        \vdots & \ddots & \vdots\\
        \hat{z}_{N,1} & \dots & \hat{z}_{N, T}
    \end{pmatrix}
    \in
    \mathbb{R}^{N \times T}
    ,
\end{align}
where $\hat{z}_{i, t} = \Phi^{-1}(\hat{y}_{i,t})$.
Finally, let $\hat{\boldsymbol{\Sigma}} \in \mathbb{R}^{N \times N}$ be the Pearson correlation matrix of the $\hat{z}$ variables, estimated as
\begin{align}
    \hat{\boldsymbol{\Sigma}} = \frac{1}{T} \hat{\mathbf{Z}} \, \hat{\mathbf{Z}}^{\top},
\end{align}
which defines a Gaussian copula function $\mathbf{C}_{\hat{\boldsymbol{\Sigma}}}$.
Note that $\hat{\boldsymbol{\Sigma}}$ is estimated using only historical data, and that it captures dependencies across site-level forecasts.

\subsection{Aggregating Site-level Probabilistic Forecasts}
\label{sec:methodology:aggregation}

Now assume the copula function $\mathbf{C}_{\hat{\boldsymbol{\Sigma}}}$ is known.
Let $\tau > T$ denote a previously-unseen time step,
and assume site-level probabilistic forecasts $(\hat{F}_{i, \tau})_{i \in \mathcal{N}}$ are available.
The proposed probabilistic forecast aggregation strategy combines
the marginal forecasts with the learned copula function,
to model the joint distribution

\begin{align}
    \hat{\mathbf{F}}_{\tau}(x_{1, \tau}, ..., x_{N, \tau})
    &= \mathbf{C}_{\hat{\boldsymbol{\Sigma}}} \big(
        \hat{F}_{1, \tau}(x_{1, \tau}), ..., \hat{F}_{N, \tau}(x_{N, \tau})
    \big),
\end{align}

from which the aggregated distribution $\hat{F}_{0, \tau}$ can be derived.

In practice, it may not be possible to derive $\hat{F}_{0, \tau}$ in analytical form.
Therefore, the paper uses a sampling-based Monte Carlo approach to approximate $\hat{F}_{0, \tau}$
in a non-parametric way.
Given a desired number of samples $S$,
the method starts by sampling $S$ points from a multi-variate normal distribution
$\mathcal{MVN}(0, \hat{\boldsymbol{\Sigma}})$, denoted by 
\begin{align}
    \tilde{\mathbf{Z}}
    &=
    \begin{pmatrix}
        \tilde{z}_{1, \tau}^{(1)} & \dots & \tilde{z}_{1, \tau}^{(S)}\\
        \vdots & \ddots & \vdots\\
        \tilde{z}_{N, \tau}^{(1)} & \dots & \tilde{z}_{N, \tau}^{(S)}
    \end{pmatrix}
    \in \mathbb{R}^{N \times S},
\end{align}
where $\tilde{z}_{i, \tau}^{(s)}$ denotes the $i$-th coordinate of the $s$-th sample.
Then, define 
\begin{align}
    \tilde{\mathbf{Y}} 
    = 
    \begin{pmatrix}
        \tilde{y}_{1, \tau}^{(1)} & \dots & \tilde{y}_{1, \tau}^{(S)}\\
        \vdots & \ddots & \vdots\\
        \tilde{y}_{N, \tau}^{(1)} & \dots & \tilde{y}_{N, \tau}^{(S)}
    \end{pmatrix}
    =
    \Phi(\tilde{\mathbf{Z}})
    =
    \begin{pmatrix}
        \Phi(\tilde{z}_{1, \tau}^{(1)}) & \dots & \Phi(\tilde{z}_{1, \tau}^{(S)})\\
        \vdots & \ddots & \vdots\\
        \Phi(\tilde{z}_{N, \tau}^{(1)}) & \dots & \Phi(\tilde{z}_{N, \tau}^{(S)})
    \end{pmatrix}
    \in \mathbb{R}^{N \times S}.
\end{align}
\emph{Correlated}, site-level generation samples are then recovered as 
\begin{align}
    \tilde{\mathbf{X}}
    =
    \begin{pmatrix}
        \hat{F}_{1,\tau}^{-1}(\tilde{y}_{1, \tau}^{(1)}) & \dots & \hat{F}_{1,\tau}^{-1}(\tilde{y}_{1, \tau}^{S})\\
        \vdots & \ddots & \vdots\\
        \hat{F}_{N,\tau}^{-1}(\tilde{y}_{N, \tau}^{(1)}) & \dots & \hat{F}_{N,\tau}^{-1}(\tilde{y}_{N, \tau}^{(S)})
    \end{pmatrix}
    \in \mathbb{R}^{N \times S}.
\end{align}
Note that $\tilde{\mathbf{X}}$ samples follow the marginal distributions specified by marginal forecasts $(\hat{F}_{i, \tau})_{i \in \mathcal{N}}$, but are correlated through the copula function $\mathbf{C}_{\hat{\boldsymbol{\Sigma}}}$.
Finally, fleet-level samples are obtained through the summation
\begin{align}
    \forall s \in \{1, ..., S\}, \tilde{x}_{0, \tau}^{(s)} = \sum_{i=1}^{N} \tilde{x}_{i, \tau}^{(s)},
\end{align}
which yields a sample-based representation of $\hat{F}_{0, \tau}$.
This approach has the advantage of supporting parametric and non-parametric
representations of the marginal forecasts, which are often provided in the form of quantiles.

\section{Numerical Experiments}
\label{sec:experiments}

This section presents a numerical study of the proposed approach, on a publicly-available dataset provided by the National Renewable Energy Laboratory (NREL) \cite{NREL_PERFORMDataset}. 
The experiments consider day-ahead forecasting of solar power generation across 751 sites across the on the Midcontinent Independent System Operator (MISO) footprint.

\subsection{Experimental Setting}

The dataset contains two primary features. The first is \emph{Actuals}, which represent the power generation at each site over time. 
This data is provided at the site level 

with a temporal resolution of five minutes. 
The second feature is \emph{Marginal Forecasts}, provided as 99 quantile values corresponding to percentile-based quantile levels.
The paper considers day-ahead forecasts, which are generated daily with an 11-hour lead time, and cover a 24-hour horizon (midnight to midnight of the subsequent day) with an hourly resolution.
Forecasts are provided for the entire year 2019, hence 12 months of data is considered for training and evaluation. 
Records corresponding to nighttime hours and near sunset and sunrise, where the generation capacity was below 4\%, were excluded to ensure data quality.

In addition to the proposed \emph{Copula} method, this study evaluates two baseline approaches.
The first baseline, denoted by \emph{Indep}, considers marginal forecasts as independent, i.e., it ignores the correlation structure.
This approach is implemented by generating $S$ samples from marginal forecasts independently, then aggregating the samples.
The second baseline, denoted by \emph{Q-sum}, assumes that marginal forecasts are provided in the form of $K$ quantiles for each site.
It then generates a fleet-level distribution by aggregating individual quantiles directly.
For instance, in \emph{Q-sum}, the 75\% quantile of total generation is obtained by summing the 75\% quantile of each marginal forecast.

All methods are evaluated on the NREL dataset described above.
The dataset is split between training (11 months) and test (1 month) data.
The paper measures the quality of the aggregated forecasts using the Prediction Interval Coverage Probability (PICP) and Average Interval Width (AIW) metrics.
Given $\alpha \in [0, 1]$, well-calibrated forecasts should produce $1-\alpha$ prediction intervals whose PICP equals $1 - \alpha$.
PICP values higher than the target coverage $1-\alpha$ indicate that forecasts are \emph{over-covering}, which typically indicates that prediction intervals are too large.
On the other hand, smaller-than-target PICP values are an indication of \emph{under coverage}.
Naturally, larger prediction intervals yield higher PICP.
Hence, one should consider PICP and AIW metrics jointly.
Namely, for a given coverage level, prediction intervals should be as small as possible, i.e., the AIW metric should be small.

\subsection{Results}
\label{sec:experiments:results}

\begin{table}[!t]
    \centering
    \caption{Comparison of aggregate day-ahead solar forecasts quality.}
    \label{tab:SD}
    \begin{tabular}{lrrrrrrrr}
        \toprule
            & \multicolumn{2}{c}{\small $(1{-}\alpha) = 90\%$}
            & \multicolumn{2}{c}{\small $(1{-}\alpha) = 80\%$}
            & \multicolumn{2}{c}{\small $(1{-}\alpha) = 70\%$}
            & \multicolumn{2}{c}{\small $(1{-}\alpha) = 60\%$} \\
        \cmidrule(lr){2-3}
        \cmidrule(lr){4-5}
        \cmidrule(lr){6-7}
        \cmidrule(lr){8-9}
        Method
            & PICP $\uparrow$ & AIW $\downarrow$
            & PICP $\uparrow$ & AIW $\downarrow$
            & PICP $\uparrow$ & AIW $\downarrow$
            & PICP $\uparrow$ & AIW $\downarrow$\\
        \midrule
        Indep 
            & 13.7 &   874.0 
            & 10.9 &   675.7 
            &  9.3 &   544.0 
            &  7.3 &   440.3 \\
        Q-sum 
            & 97.2 & 14727.2 
            & 90.3 & 11624.7 
            & 85.9 &  9094.1 
            & 80.2 &  7404.4 \\
        Copula 
            &  78.2 & 7437.5 
            &  69.7 & 5769.7
            &  59.7 & 4731.4 
            &  50.4 & 3849.1 \\
       
        \bottomrule
    \end{tabular}
\end{table}

\begin{figure}[!t]
    \centering
    \includegraphics[width=\linewidth]{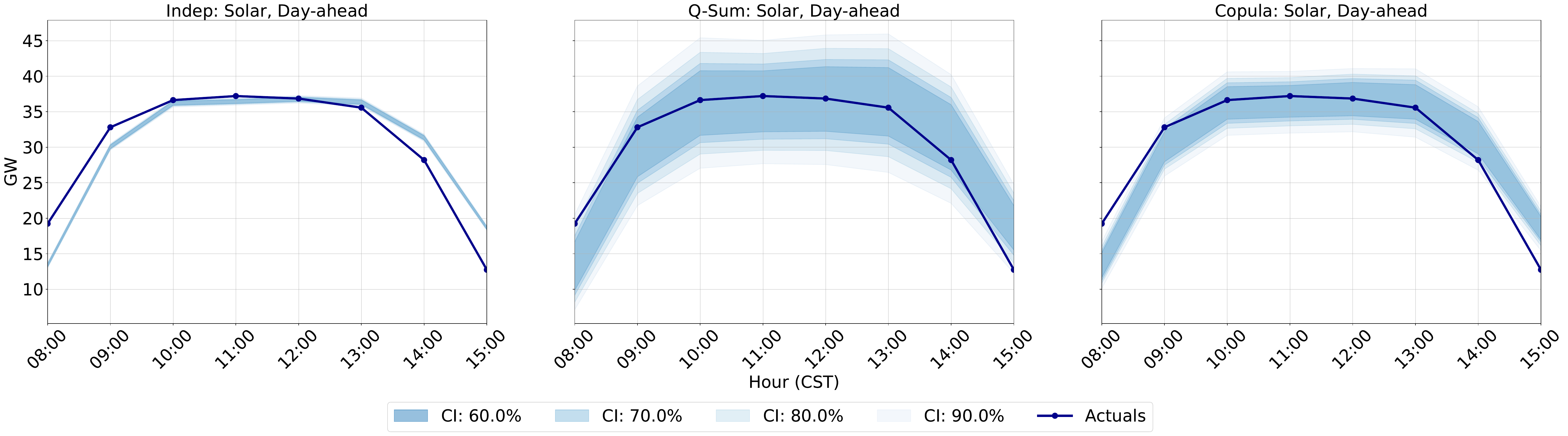}
    \caption{Actual total solar generation and prediction intervals from aggregated day-ahead forecasts. Left: \emph{Indep}; Middle: \emph{Q-sum}; Right: \emph{Copula}.
    Sunrise and sunset times are approximately  7:20am and 5pm, respectively.}
    \label{fig:Copula-hourly}
    \vspace{-0.5em}
\end{figure}

Table \ref{tab:SD} presents the PICP and AIW for the proposed method and baselines across four confidence levels, ranging from $1-\alpha = 60\%$ to $1-\alpha = 90\%$. 
In addition, Figure \ref{fig:Copula-hourly} displays the actual total generation and prediction intervals produced by each method on Dec 5th, 2019.

A first observation is that the naive \emph{Indep} approach performs poorly, as demonstrated by its very low PICP scores.
This behavior is expected because \emph{Indep} considers marginal (site-level) forecasts to be independent.
Thereby, the aggregate forecast error is the sum of $N$ independent site-level forecast errors, hence, per the central limit theorem, its variance is small (relative to the mean).
Therefore, as illustrated by Figure \ref{fig:Copula-hourly}, \emph{Indep} yields prediction intervals with small width (about 10x smaller than \emph{Copula}) and, hence, poor coverage.

In contrast, the \emph{Q-sum} method achieves high PICP but at the cost of an excessively large AIW, making it impractical for real-world applications.
Also note that the PICP of \emph{Q-sum} is substantially higher than target coverage, especially for lower confidence levels: the 60\% prediction intervals produced by \emph{Q-sum} yield a PICP of 80\%.
These results suggest that \emph{Q-sum} does not capture aggregate generation well.
Finally, the proposed \emph{Copula} method displays PICP scores that are consistently about 10\% lower than target coverage.
This difference is explained by the fact that forecasts consistently miss actual generation during the last hour before sunset, thus resulting in lower PICP.
The prediction intervals produced by \emph{Copula} are roughly 2x smaller than those produced by \emph{Q-sum}.

\begin{figure}[!t]
    \centering
    \includegraphics[width=0.45\linewidth]{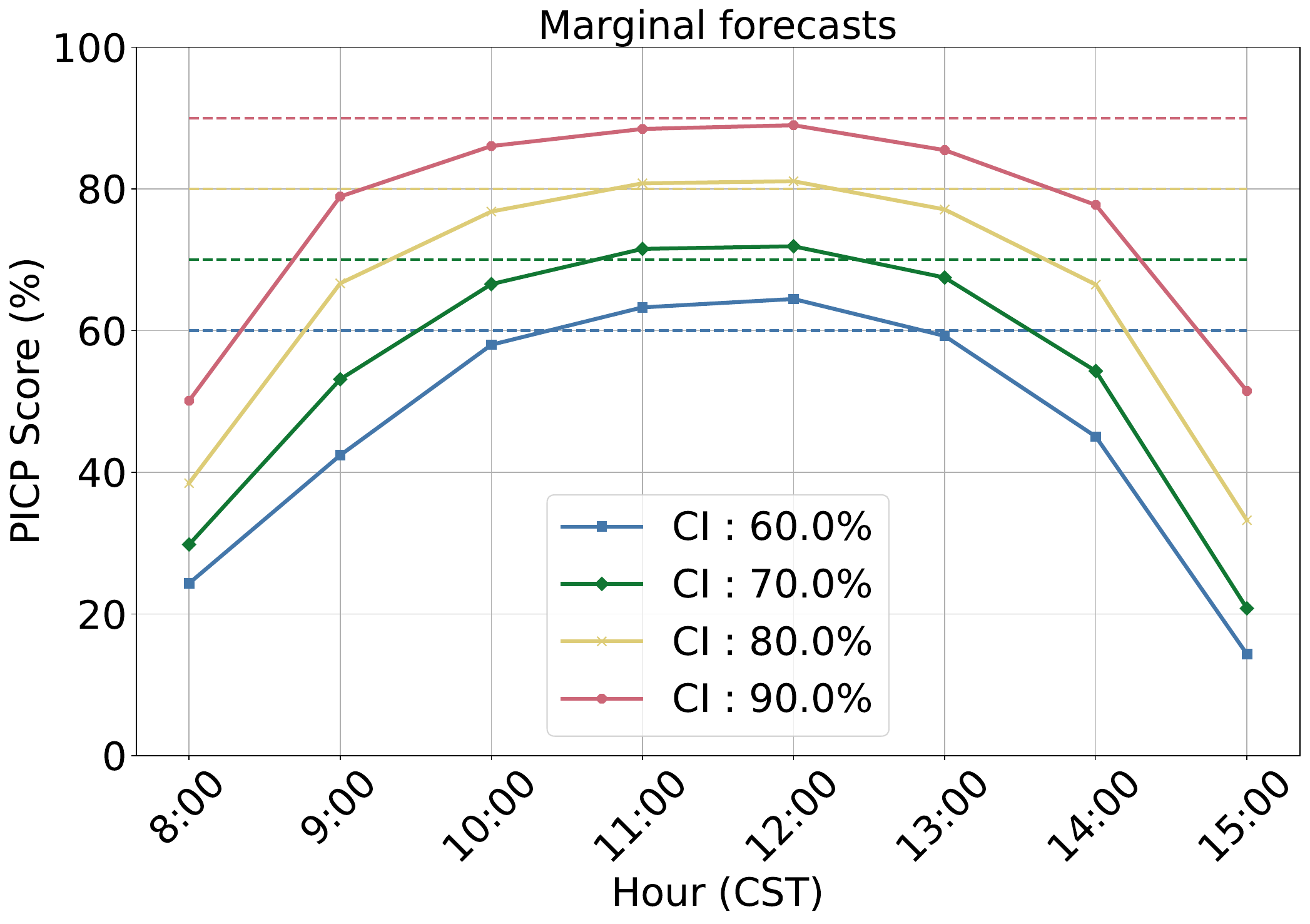}
    \caption{The PICP accuracy of sites marginal forecasts}
    \label{fig:marg}
    \vspace{-0.5em}
\end{figure}

It is important to note that the quality of the probabilistic aggregated forecasts is highly dependent on the accuracy of individual marginal forecasts.
To that end, Figure \ref{fig:marg} illustrates the average PICP of marginal forecasts across individual sites. 
While the PICP aligns with target coverage levels during peak hours (10:00–13:00), there is notable under-coverage during early morning and late evening hours.
This explains the observed under-coverage of the \emph{Copula} method in Table \ref{tab:SD}.
Note that, reciprocally, over-covering marginal forecasts would typically yield over-coverage at the aggregate level.

Finally, while maintaining an overall high coverage rate is essential, it is particularly important to ensure reliable coverage during peak power generation hours.
Indeed, operational decisions are less impacted by forecast errors during during early morning and late afternoon hours, when overall generation is low.
Accordingly, Figure \ref{fig:Copula-hourly} presents the average PICP for each method across different hours of the day. 
Consistent with previous observations, the \emph{Indep} method performs poorly throughout the day, with the PICP of 90\% prediction intervals never exceeding 30\%. 
Next, \emph{Q-sum} displays 100\% coverage at all confidence levels between 9am and 1pm.
This is not desirable, since lower confidence levels should result in smaller prediction intervals.
In contrast, the PICP of the proposed \emph{Copula} varies across the day, with peak hours displaying higher coverage.
Note, that, as mentioned earlier, PICP for \emph{Copula} is close to zero at 3pm, which explains the results of Table \ref{tab:SD}.

\begin{figure}[!t]
    \centering
    \includegraphics[width=\linewidth]{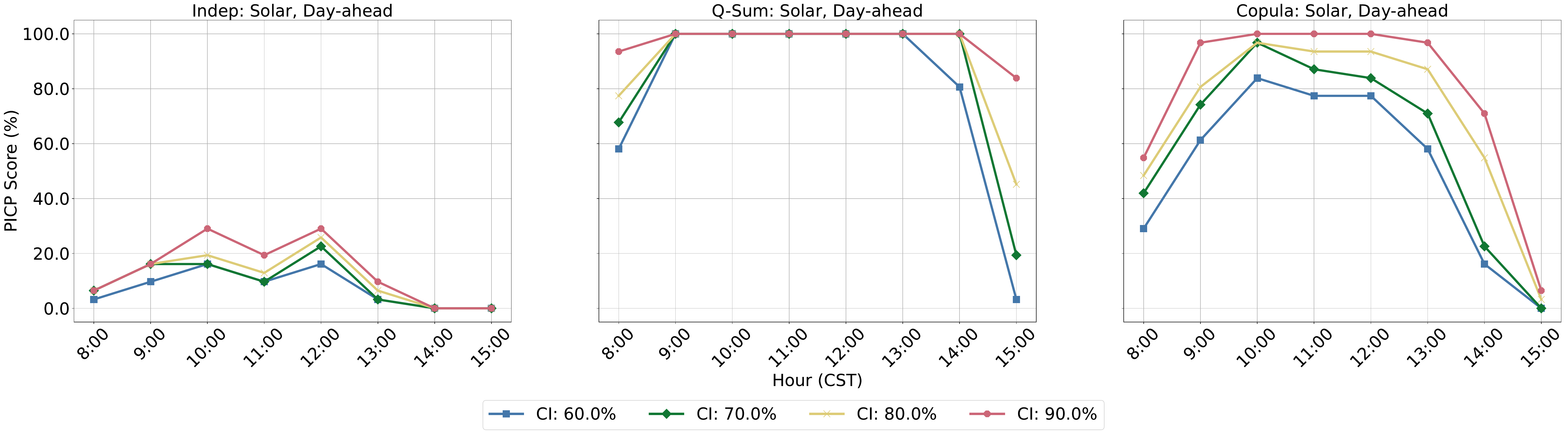}
    \caption{Average PICP scores across each hour of the day.
    Left: \emph{Indep};
    Middle: \emph{Q-sum};
    Right: \emph{Copula}}
    \label{fig:Copula-daily}
    \vspace{-0.5em}
\end{figure}

\section{Conclusion}
\label{sec:conclusion}

This paper proposed a copula-based method for probabilistic aggregation in power systems, designed to capture dependencies across different sites and provide realistic probabilistic forecasts at the fleet level. 
The results demonstrated that the proposed approach achieves a high coverage rate while maintaining a reasonable prediction interval width. 
Notably, the method ensures reliable coverage during peak generation hours, which is crucial for informed decision-making in operational settings.

Additionally, this study emphasized the significance of accurate marginal forecasts in the aggregation process, as their quality directly impacts overall performance.
Future research will explore extending this approach to wind power generation and developing multi-level aggregation frameworks to enhance scalability and applicability across diverse energy systems.

\section*{Acknowledgments}
This research was partly funded by NSF award 2112533 and ARPA-E PERFORM award AR0001136.

\bibliographystyle{unsrt}  
\bibliography{references}

\begin{thebibliography}{10}

\bibitem{Zhang2025_ProbabilisticForecasting}
Hanyu Zhang, Reza Zandehshahvar, Mathieu Tanneau, and Pascal {Van Hentenryck}.
\newblock Weather-informed probabilistic forecasting and scenario generation in power systems.
\newblock {\em Applied Energy}, 384:125369, 2025.

\bibitem{GraphCast}
Remi~Lam et~al.
\newblock Learning skillful medium-range global weather forecasting.
\newblock {\em Science}, 382(6677):1416--1421, 2023.

\bibitem{miswan2016arima}
Nor~Hamizah Miswan, Rahaini~Mohd Said, and Siti Haryanti~Hairol Anuar.
\newblock Arima with regression model in modelling electricity load demand.
\newblock {\em Journal of Telecommunication, Electronic and Computer Engineering (JTEC)}, 8(12):113--116, 2016.

\bibitem{alberg2018short}
Dima Alberg and Mark Last.
\newblock Short-term load forecasting in smart meters with sliding window-based arima algorithms.
\newblock {\em Vietnam Journal of Computer Science}, 5:241--249, 2018.

\bibitem{zahid2019electricity}
Maheen Zahid, Fahad Ahmed, Nadeem Javaid, Raza~Abid Abbasi, Hafiza~Syeda Zainab~Kazmi, Atia Javaid, Muhammad Bilal, Mariam Akbar, and Manzoor Ilahi.
\newblock Electricity price and load forecasting using enhanced convolutional neural network and enhanced support vector regression in smart grids.
\newblock {\em Electronics}, 8(2):122, 2019.

\bibitem{8272993}
Ali Lahouar, Amal Mejri, and Jaleleddine Ben~Hadj Slama.
\newblock Probabilistic day-ahead load forecast using quantile regression forests.
\newblock In {\em 2017 International Conference on Engineering \& MIS (ICEMIS)}, pages 1--6, 2017.

\bibitem{he2020day}
Feifei He, Jianzhong Zhou, Li~Mo, Kuaile Feng, Guangbiao Liu, and Zhongzheng He.
\newblock Day-ahead short-term load probability density forecasting method with a decomposition-based quantile regression forest.
\newblock {\em Applied Energy}, 262:114396, 2020.

\bibitem{8863951}
Zhaojing Cao, Can Wan, Zijun Zhang, Furong Li, and Yonghua Song.
\newblock Hybrid ensemble deep learning for deterministic and probabilistic low-voltage load forecasting.
\newblock {\em IEEE Transactions on Power Systems}, 35(3):1881--1897, 2020.

\bibitem{amarasinghe2017deep}
Kasun Amarasinghe, Daniel~L Marino, and Milos Manic.
\newblock Deep neural networks for energy load forecasting.
\newblock In {\em 2017 IEEE 26th international symposium on industrial electronics (ISIE)}, pages 1483--1488. IEEE, 2017.

\bibitem{taieb2021hierarchical}
Souhaib~Ben Taieb, James~W Taylor, and Rob~J Hyndman.
\newblock Hierarchical probabilistic forecasting of electricity demand with smart meter data.
\newblock {\em Journal of the American Statistical Association}, 116(533):27--43, 2021.

\bibitem{sun2019conditional}
Mucun Sun, Cong Feng, and Jie Zhang.
\newblock Conditional aggregated probabilistic wind power forecasting based on spatio-temporal correlation.
\newblock {\em Applied Energy}, 256:113842, 2019.

\bibitem{taieb2017coherent}
Souhaib~Ben Taieb, James~W Taylor, and Rob~J Hyndman.
\newblock Coherent probabilistic forecasts for hierarchical time series.
\newblock In {\em International Conference on Machine Learning}, pages 3348--3357. PMLR, 2017.

\bibitem{NREL_PERFORMDataset}
Brian Sergi et~al.
\newblock {ARPA-E PERFORM datasets}, 2022.

\end{thebibliography}

\end{document}